\documentclass[aps,prb,superscriptaddress,twocolumn,showpacs,amsmath,amssymb]{revtex4}
\usepackage{graphicx}

\begin{document}

\title{Periodic-orbit resonance in the quasi-1D organic superconductor 
(TMTSF){$_2$}ClO{$_4$}}

\author{S. Takahashi}
\affiliation{Department of Physics, University of Florida,
Gainesville, FL 32611, USA}
\author{S. Hill} \email[corresponding
author, Email:]{hill@phys.ufl.edu} \affiliation{Department of
Physics, University of Florida, Gainesville, FL 32611, USA}
\author{S. Takasaki}
\affiliation{Department of Material Science, Graduate School of
Science, Himeji Institute of Technology, 3-2-1 Kouto,
Kamigori-cho, Ako-gun, Hyogo 678-1297, Japan}
\author{J. Yamada}
\affiliation{Department of Material Science, Graduate School of
Science, Himeji Institute of Technology, 3-2-1 Kouto,
Kamigori-cho, Ako-gun, Hyogo 678-1297, Japan}
\author{H. Anzai}
\affiliation{Department of Material Science, Graduate School of
Science, Himeji Institute of Technology, 3-2-1 Kouto,
Kamigori-cho, Ako-gun, Hyogo 678-1297, Japan}

\date{\today}

\begin{abstract}
We report the observation of periodic-orbit resonance (POR) in a
metal possessing pure quasi-one-dimensional (Q1D)
Fermiology$\--$namely, the organic linear-chain compound
(TMTSF)$_2$ClO$_4$, whose Fermi surface consists of a pair of weakly
warped sheets. The POR phenomenon is related to weak inter-chain
coupling, which allows electrons constrained on open trajectories to
acquire small transverse velocities. Application of an appropriately
oriented magnetic field induces periodic motion transverse to the
chain direction and, hence, to a resonance in the AC conductivity.
This resonance phenomenon is closely related to cyclotron resonance
observed in metals with closed Fermi surfaces. For the Q1D POR, the
field orientation dependence of the resonance is related simply to
the Fermi velocity and lattice periodicity.


\end{abstract}


\maketitle

The organic metal (TMTSF)$_2$ClO$_4$ belongs to the family of
quasi-one-dimensional (Q1D) Bechgaard
salts\cite{Bechgaard80,Ishiguro} having the common formula
(TMTSF)$_2$X; TMTSF is an abbreviation for
tetramethyl-tetraselenafulvalene, and the anion X is ClO$_4$,
PF$_6$, ReO$_4$, etc. For a long time, (TMTSF)$_2$X was considered
to be a conventional superconductor with a low T$_c$ ($\sim 1$~K).
However, in the last few years, there have appeared clear evidences
indicating an unconventional (triplet) character to the
superconductivity in both the X~$=$~ClO$_4$ and PF$_6$
salts.\cite{Lee95,LeePRB00,LeePRL02} For example, Lee et al. have
shown that the upper critical field ($H_{c2}$) diverges as
T$\rightarrow0$ for fields applied along certain
directions.\cite{Lee95,LeePRB00} This divergence is most pronounced
in the PF$_6$ salt (under pressure), with $H_{c2}$ exceeding the
Pauli paramagnetic limit by up to a factor of four.\cite{LeePRB00}
Furthermore, $^{77}$Se NMR studies reveal no Knight shift upon
cooling through the superconducting transition, again suggesting
spin-triplet pairing.\cite{LeePRL02}


Recent theoretical studies\cite{LouatiPRB00,TanumaPRB02,Cherng03}
have shown that the superconductivity in Q1D systems may depend
strongly on the nesting properties of the Fermi surface (FS), as
well as on the conduction bandwidth. Therefore, it is important to
be able to make detailed measurements of their band structures and
FS topologies. The FS of (TMTSF)$_2$ClO$_4$ reflects its crystal
structure, which has a highly anisotropic character.\cite{Ishiguro}
The planar TMTSF cations stack face-to-face in columns along the
crystallographic $a$-axis. These columns then form layers parallel
to the $ab$-plane. The strongest overlap between the partially
occupied $\pi$-orbitals on the TMTSF molecules occurs along the
columns. Therefore, the conductivity is highest (bandwidth is
greatest) along the $a$-axis. The orbital overlap between the TMTSF
columns, {\em i.e.} within the layers (intermediate direction), is
about 10 times less, while the bandwidth is considerably weaker in
the interlayer direction.\cite{Kikuchi82,Grant83} The electronic
band structure may be calculated using a tight binding
approximation. The anisotropy of the tight binding transfer
integrals ($t_a:t_b:t_c$) is about 200:20:1 meV. As a result, the FS
consists of a pair of weakly warped sheets,\cite{Grant83} oriented
perpendicular to the $a$-axis (see Fig.~1), i.e. the FS may be
regarded as Q1D.

DC magnetoresistance measurements have proven to be very useful in
determining the FS topologies of low-dimensional organic
conductors.\cite{Ishiguro,OsadaPRB92,DannerPRL94,ChashechPRB97,Chashech98,LeePRB98,Kobayashi02,Kang03}
In particular, a number of interesting effects are observed when
measurements are made as a function of magnetic field orientation
$-$ so-called Angle-dependent Magneto-Resistance Oscillations
(AMRO). As we have demonstrated in recent studies of numerous
quasi-two-dimensional (Q2D) organic conductors, additional important
information may be obtained from high frequency (microwave) AMRO
studies.\cite{HillPRB97,KovalevPRB02,KovalevPRL} High frequency
implies $\omega\tau>1$ ($\tau^{-1}\equiv$~scattering rate), and that
$2\pi/\omega$ is comparable to the period of any electronic motion
caused by the application of a magnetic field (see below). Here, we
report similar measurements for the pure Q1D (TMTSF)$_2$ClO$_4$
system. Semiclassical calculations show that a special kind of Q1D
cyclotron resonance (CR), or periodic-orbit resonance
(POR\cite{HillPRB97}), should be
observed.\cite{HillPRB97,KovalevPRB02,KovalevPRL,BlundellPRB97,ArdavanPRL98}
Indeed, this behavior is closely related to the semiclassical Q1D
AMRO first discussed by Osada.\cite{OsadaPRB92} Using this effect,
one can determine the Fermi velocity ($v_F$) for (TMTSF)$_2$ClO$_4$
without dependence on any parameters other than the lattice
periodicity. The idea of cyclotron orbits in a Q1D system should not
be taken literally. Nevertheless, {\em it is} the Lorentz force
which causes the periodic electronic motion which ultimately gives
rise to this unusual POR phenomenon, as we now briefly describe with
the aid of Fig.~1. We note that our discussion of the POR is based
on the semiclassical Boltzmann transport equation.\cite{OsadaPRB92}
We discuss the validity of this approach later in this article.

Due to the finite bandwidth (finite orbital overlap) in the
intermediate conductivity direction, the FS sheets become warped, as
represented by the corrugations in Fig.~1. This warping may be
decomposed into Fourier harmonics, each characterized by some vector
{$\bf R$} in real space. In fact, one expects a different warping
vector for every finite transfer integral. Thus, in general, one
requires a set of indices ($m,n$) in order to keep track of all of
the warping harmonics {\bf R}$_{m,n}$. However, due to the
simplicity of the experimental data, we restrict the following
discussion to a single vector {\bf R}$_\parallel$ (Fig.~1),
representing the projection of a single warping vector onto the
plane of the FS. In the presence of a magnetic field ({\em B}), an
electron moves along the FS with a constant rate of change of
momentum ($\hbar dk/dt$) in a plane perpendicular to the magnetic
field, and with its velocity always directed perpendicular to the
FS. Thus, its motion is mainly directed along the $a$-axis
($v_\parallel$). However, due to the periodic FS corrugations, the
electron's transverse velocity ($v_\perp$) oscillates slightly about
zero (see trajectory in Fig.~1). In general, for a three dimensional
FS, this effect gives rise to periodic oscillations of both velocity
components within the plane of the Q1D FS ($b'c^*$-plane in this
case). The frequency of these oscillations is proportional to (i)
the magnitude of $\bf R_\parallel$, and (ii) the magnetic field
component perpendicular to the warping vector ({\bf R}$_\parallel$)
and parallel to the FS. Each harmonic of the warping, therefore,
gives rise to a distinct resonance in the $b'c^*$-plane microwave
conductivity, with a frequency that scales with $B$, as is also the
case for conventional CR. Meanwhile, each Q1D resonance has a
distinct angle dependence characterized by the warping vector $\bf
R_\parallel$. However, instead of the CR mass, the resonance
frequency $\nu$ depends on the Q1D Fermi velocity ($v_F$) though the
following equations:\cite{BlundellPRB97,KovalevPRB02}

\smallskip
\begin{subequations}
\label{rescond}
\begin{equation}
\frac{\nu}{B_{res}}=\frac{e v_F R_{\parallel}}{h}\left|
{\sin(\theta_{b'c^*}-\theta_\circ)} \right|,
\end{equation}
\begin{equation}
\frac{\nu}{B_{res}}=\frac{e v_F
R_{\parallel}}{h}\left|\cos(\theta_\perp)
{\sin(\theta_\circ^\prime)} \right|.
\end{equation}
\end{subequations}
\

{\noindent{Eq.~1(a) corresponds to magnetic field rotations parallel
to the plane of the FS; $\theta_{b'c^*}$ is the angle between $c^*$
and the projection of {\em B} onto the $b^\prime c^*$-plane; and
$\theta_\circ$ represents the angle between $c^*$ and {\bf
R}$_\parallel$. Eq.~1(b) corresponds to magnetic field rotations
away from the $b^\prime c^*$-plane, towards the $a$-axis (i.e. in a
plane $\perp$ to the FS); $\theta_\perp$ is the angle between the
$b^\prime c^*$-plane and {\em B}; and $\theta_\circ^\prime$
represents the angle between the plane of rotation and {\bf
R}$_\parallel$. Finally, in both cases, $B_{res}$ is the resonance
field. These geometries are depicted in Fig.~1. It is the
conductivity parallel to the Q1D Fermi surface ($b'c^*$-plane) which
has a maximum at the resonance, and the amplitude of the resonance
is proportional to the amplitude of the relevant Fourier component
($G_{m,n}$) of the FS corrugation. Thus, angle dependent
measurements enable a determination of the orientations of warping
vectors (or transfer integrals). A more general expression for a
material possessing a Q1D FS with many warping components is given
in ref.~[\onlinecite{KovalevPRB02}]. While similar information may
also be deduced from DC AMRO ($\nu \sim 0$), one obtains only the
directions of warping vectors from such studies, i.e. AMRO are
observed at the zeros of Eq.~1(a), where $\theta_{b'c^*} =
\theta_\circ$. On the other hand, finite frequency Q1D POR
additionally enables a determination of the product $v_FR_\parallel$
(see Eq.~1). Since the warping vectors ({\bf R}$_{m,n}$) are usually
related to the underlying lattice constants, which are well known
from X-ray data, Q1D POR provides the most direct means of measuring
$v_F$ in Q1D metals. Until now, this technique has been limited to
systems with strongly warped open FS sections (i.e. Q2D systems),
leading to a strong harmonic content of the
POR.\cite{ArdavanPRL98,KovalevPRB02,Oshima} The present study is the
first of its kind for a member of the widely studied Q1D
(TMTSF)$_2$X family. We note that a previous investigation within
the field-induced-spin-density-wave (FISDW) phase of
(TMTSF)$_2$ClO$_4$ ($B>6$~T, T$\sim 1.2$~K) tentatively ascribed
several dips in the far-infrared magneto-reflectivity to
inter-Landau-level (i.e. CR-like) transitions across the FISDW
gap.\cite{Perel}


Microwave measurements were carried out using a millimeter-wave
vector network analyzer and a high sensitivity cavity perturbation
technique; this instrumentation is described in detail
elsewhere.\cite{MolaRSI} In order to enable in-situ rotation of the
sample relative to the applied magnetic field, we employed two
methods. The first involved a split-pair magnet with a 7~T
horizontal field and a vertical access. Smooth rotation of the
entire rigid microwave probe, relative to the fixed field, was
achieved via a room temperature stepper motor (with $0.1^\circ$
resolution). The second method involved in-situ rotation of the
end-plate of a cylindrical cavity, mounted with its axis transverse
to a 17~T superconducting solenoid. Details concerning this cavity,
which provides an angle resolution of $0.18^\circ$, have been
published elsewhere.\cite{takahashi04} Several needle shaped samples
were separately investigated by placing them in one of two
geometries within the cylindrical TE011 cavities, enabling (i) field
rotation in the $b'c^*$-plane, and (ii) rotation away from the
$b^\prime c^*$-plane towards the $a$-axis. Each sample was slowly
cooled ($< 0.1$~K/min between 32~K and 17~K) through the anion
ordering transition at 24~K to obtain the low-temperature metallic
state.\cite{Ishiguro} All experiments were performed at 2.5~K, and
data for two of the samples are presented in this paper (labeled A
and B, dimensions $\sim 1\times0.2\times0.1$~mm$^3$). In order to
verify that the relaxed state is reproducibly achieved, we can
observe signatures of the FISDW phase transition in the microwave
response (see Fig.~2) for experiments conducted in the 17~T magnet.
Studies of the angle dependence of the FISDW transition also allow
us to determine the orientations of the $b'$ and $c^*$ directions
in-situ, i.e. simultaneous to the POR measurements.


In Fig.~2, we present 61.8~GHz absorption data for sample~A (2nd
cooling) in fields up to 15~T. The magnetic field was rotated in the
$b^\prime c^*$-plane ($\perp$ needle axis), and successive traces
were taken in $8.8^\circ$ steps; the angle ($\theta_{b^\prime c^*}$)
refers to the field orientation relative to the $c^*$ axis. Two
pronounced angle-dependent features are apparent in the data. At the
lowest fields, a peak in absorption is observed (labeled POR) which
moves to higher magnetic fields upon rotating the field away from
the $b^\prime$ direction ($\theta_{b^\prime c^*}=90^\circ$). It is
this peak which corresponds to the POR. Its position in field is
proportional to the frequency (not shown), as expected for a
cyclotron-like resonance. At higher fields (above $\sim9$~T), a
sharp kink in the absorption is observed (labeled $B_{\rm FISDW}$),
which moves to higher fields upon rotating the field away from the
$c^*$ direction ($\theta_{b^\prime c^*}=0^\circ$). The position of
this feature does not depend on the measurement frequency. In fact,
its angle and temperature dependence are in excellent agreement with
the expected behavior of the $N=0$ FISDW phase
boundary.\cite{Ishiguro,Osada01} We are thus able to use the
angle-dependence of this kink to determine the field orientation
relative to the $b'$ and $c^*$ directions. Qualitatively similar POR
data were obtained for this (1st cooling) and other samples in the
lower field magnet, both for $b'c^*$-plane field rotations and for
rotations away from the $b'c^*$-plane (see Fig.~3).

For all measurements reported in this paper, the sample was
positioned within the cavity in such a manner so as to excite
currents in the low conductivity $b'c^*$-plane, as required for Q1D
POR. In fact, it is very difficult to measure the $a$-axis
conductivity using the cavity perturbation technique. The
penetration depths for the typical frequencies used in this study
are of order $100~\mu$m for currents along $c^*$, and $5~\mu$m for
currents along $b^\prime$. Consequently, the electromagnetic fields
penetrate well into the sample, thus accounting for the
approximately Lorentzian CR lineshape which is characteristic of the
bulk conductivity.\cite{HillPRB00} This rules out any possibility
that the observed resonance could be due to the Azbel'-Kaner-type CR
originally predicted for Q1D conductors by Gor'kov and
Lebed.\cite{GorkovPRL93} In particular, one would expect to observe
several CR harmonics in such a case, but only for the surface
resistance parallel to $a$. Furthermore, it is unlikely that the
conditions for the Gor'kov and Lebed CR could ever be satisfied in
(TMTSF)$_2$ClO$_4$, namely that the $b'$-axis penetration depth
($\sim 3-5~\mu$m) be considerably less than the amplitude of the
$b'$-axis motion ($\sim  0.1~\mu$m at 2~tesla) caused by the FS
corrugations (see Fig.~1) and, simultaneously, $\omega_c\tau >> 1$.

In Fig.~3a we plot the $b'c^*$-plane angle dependence of the ratio
$\nu /B_{res}$ for two samples (A and B), and for several
measurement frequencies. $B_{res}$ is determined simply from the
position of maximum absorption due to the POR (low field dashed
curve in Fig.~2). The POR data (open symbols) agree extremely well
with Eq.~1a (solid curves), with relatively little scatter in the
data for the various measurements. Also displayed in Fig.~3a is the
angle-dependence of the FISDW transition field, $B_{\rm FISDW}$
(solid squares). These data scale with the inverse of the cosine of
the angle $\theta_{b^\prime c^*}$ (dashed curve), reflecting the 2D
nature of the standard theory of the FISDW.\cite{Ishiguro,Osada01}
From the $1/\cos\theta_{b^\prime c^*}$ fits to $B_{\rm FISDW}$, we
see that the zeros in $B_{res}$ (equivalent to $m/n = p/q = 0$ AMRO
minima) occur when the field is parallel to the $c^\prime$ direction
($\sim5^\circ$ away from $c^*$), in excellent agreement with DC AMRO
measurements.\cite{LeePRB98,Chashech98,Kang03,Osada91,Naughton91} In
Fig.~3b we plot the ratio $\nu /B_{res}$ for sample A, for rotation
away from the $b^\prime c^*$ plane. In this case, the zeros in $\nu
/B_{res}$ were found to lie exactly along the $a$-axis of the
crystal, as expected from Eq.~1b. The reason for the discrepancy
between the maximum value of $\nu /B_{res}$ in Figs.~3a and~3b is
due to the fact that the orientation of the $c^*$-axis was not well
known when mounting the sample for the rotations away from the
$b^\prime c^*$ plane, resulting in a mis-alignment between the plane
of rotation and $c^*$, i.e. a finite $\theta_\circ^\prime$ (Eq.~1b),
thus diminishing the maximum in $\nu /B_{res}$.

It is important to stress that the POR phenomenon (semiclassical
AMRO) is entirely expected if one assumes a 3D band structure along
with the accepted 3D hopping matrix elements. This idea was first
discussed for the case of the Bechgaard salts by
Osada,\cite{OsadaPRB92} and has subsequently been adapted by many
authors to account for both DC and high-frequency conductivity
resonances observed in several Q2D BEDT-TTF
salts.\cite{KovalevPRB02,KovalevPRL,
BlundellPRB97,ArdavanPRL98,Blundell96} However, one has to consider
whether a 3D band transport picture is appropriate for the Bechgaard
salts in view of their extreme low-dimensionality. Indeed, Lebed's
original explanation for the so-called ``magic-angle effects" (or
``Lebed resonances") involves a magnetic-field-induced dimensional
crossover. Application of a magnetic field perpendicular to the
least conducting $c^*$ direction has the effect of
localizing/confining quasiparticles to a single layer, resulting in
a crossover from coherent (3D band transport) to incoherent
interlayer transport at relatively weak fields. The three
dimensionality is subsequently restored whenever the applied field
direction is commensurate with any intermolecular hopping direction,
{\bf R}$_{m,n}$.\cite{Lebed86,Lebed89} These changes in effective
dimensionality are expected to have a profound effect on electronic
correlations, ultimately giving rise to the Lebed resistance
resonances. However, Lebed's magic-angles are indistinguishable from
the DC AMRO directions in Osada's
theory.\cite{OsadaPRB92,Blundell96} Consequently, a concensus has
not yet emerged concerning the explanation for the Lebed resonances
seen in (TMTSF)$_2$X.

Lebed's theory differs from Osada's in the sense that the
``magic-angles" really are magic, i.e. the underlying dimensionality
of the electronic system changes from two- to three- at these
angles. Consequently, one {\em should not} observe Lebed resonances
in an experiment such as the one reported here, where the field
orientation is fixed. In contrast, there are no ``magic-angles" in
Osada's theory. Conductivity resonances occur at frequencies and
angles given by $\omega=v_FG_{m,n}$, where the $G_{m,n}$
characterize the Fourier components of the FS
warping.\cite{OsadaPRB92} The zeros of $G_{m,n}$ determine the DC
($\omega=0$) AMRO minima. However, for $\omega \approx v_F
R_\parallel e B/\hbar$, the AMRO directions are determined instead
by the zeros of $(\omega-v_FG_{m,n})$, i.e. they shift with field
and frequency. Consequently, one {\em does} observe these resonances
by fixing the field orientation and sweeping only its magnitude.
Indeed, the fit to the resonance positions in Fig.~3 corresponds
precisely to the zeros of $(\omega-v_FG_{01})$, and the data
intersect the $\nu=0$ axis at precisely the expected angles
corresponding to the $p/q=0$ ($\equiv m/n$) Lebed resonances.

Based on the above discussion, and Fig.~3, our findings appear to
support Osada's theory. However, a surprising aspect of the data is
the lack of any significant harmonic content to the POR. In
contrast, low-temperature DC AMRO studies reveal several dips in
both the interlayer ($\parallel c^*$) and in-plane ($\parallel a$)
resistivities, corresponding to different ratios of the indices $p$
and $q$ which are commonly used to index AMRO
harmonics.\cite{Osada91,Naughton91,LeePRB98,Chashech98,Kang03}
Strong harmonics are also clearly seen, even in the POR data, for
the organic conductor
$\alpha$-(ET)$_2$KHg(SCN)$_4$;\cite{KovalevPRB02} although this
material has a Q2D band structure, its low-temperature electronic
properties are dominated by a strongly corrugated Q1D (open) FS
which results from the reconstruction of a high-temperature ($>8$~K)
Q2D FS.\cite{Blundell96} The high harmonic content of the POR
observed in $\alpha$-(ET)$_2$KHg(SCN)$_4$ is a direct of this
unusual low-temperature FS.\cite{Blundell96} In contrast,
(TMTSF)$_2$ClO$_4$ really is a Q1D conductor, i.e. one can
practically neglect higher order hopping terms
(next-nearest-neighbor, etc..) in a tight-binding description of the
$b^\prime c^\prime$-plane band dispersion. Consequently, the FS
warping will be dominated by the $G_{01}$ and $G_{10}$ Fourier
harmonics, and it is reasonable to expect higher-order POR harmonics
to be significantly weaker than the fundamental $p/q=0$ resonance.
Nevertheless, significant higher-order content has convincingly been
reported in DC AMRO experiments with ratios of $p/q$ up to
5.\cite{Naughton91} It should be noted, however, that all DC AMRO
experiments reporting high-harmonic content were performed either on
the X~=~PF$_6$ salt (under pressure), and/or at considerably lower
temperatures ($<1$~K) and higher fields than the present
investigation.\cite{Osada91,Naughton91,LeePRB98,Chashech98,Kang03}
In fact, there is a very limited amount of published DC AMRO data
for (TMTSF)$_2$ClO$_4$, obtained at comparable fields and
temperatures to the present investigation (T$=2.5$~K and
$B\sin\theta<3$~T), and those that can be found reveal no evidence
for higher order harmonics, i.e. only the $p/q=0$ resonance is seen.
Comparisons are also complicated by the fact that our experiments
are performed at fixed field orientations, whereas DC AMRO are
necessarily observed by rotating the magnetic field direction.
Furthermore, for most of the angles in the present investigation,
the $p/q>0$ harmonics would be expected at fields well below the
$p/q=0$ fundamental resonance. Therefore, on the basis of
comparisons with published DC AMRO data, it is {\em not} surprising
that the $p/q>0$ harmonics are not observed within the parameter
space available for these experiments. Future instrument
developments will be aimed at enabling angle-dependent measurements
at lower temperatures, as well as studying the X~=~PF$_6$ salt under
pressure.

Due to the mechanics of our experiment, we can make no statements
about ``magic-angle-effects". Indeed, it is quite possible that the
Lebed resonances are missed entirely in this investigation. However,
we wish to emphasize that the observed POR are precisely the same
phenomenon as the semiclassical AMRO discussed by
Osada,\cite{OsadaPRB92} and by many other authors in subsequent
papers.\cite{KovalevPRB02,KovalevPRL,
BlundellPRB97,ArdavanPRL98,Blundell96} We now know from these
investigations that the $p/q=0$ POR resonance is not ``magic" in the
sense that its orientation varies with frequency (Fig.~3). Whether
or not the Lebed resonances are observable at microwave frequencies,
and whether this really is a ``magic-angle effect" remains to be
seen. One intriguing possibility that cannot be ruled out is that
there is an interplay between both effects, and that the magic-angle
physics dominates the DC AMRO at high fields and low temperatures,
causing strong AMRO harmonics, whereas the present microwave studies
are sensitive only to the semiclassical physics.





From an analysis of the POR line shape (Fig.~2), we estimate a
relaxation time of about 4~ps. This is close to the typical value
estimated from the field dependence of the DC magnetoresistance
(1$-$5~ps) of a relaxed sample.\cite{ChashechPRB97} Based on this,
and on the known interlayer bandwidth, we can conclude that
(TMTSF)$_2$ClO$_4$ exhibits truly coherent 3D band transport at low
fields. However, application of a field within the layers will
obviously change this picture. Nevertheless, we believe that this
lends further support to our assertion that the POR observed in this
study correspond to the high-frequency AMRO first discussed by
Osada,\cite{OsadaPRB92} and there is no need to consider alternative
models such as those developed for incoherent interlayer
transport.\cite{McKenziePRB99}


We next turn to the pre-factor A$_\circ$ ($\equiv e v_F R_\| / h$)
in Eq.~1. From these studies, we obtain a single value of
$24(1)$~GHz/T. However, earlier studies on older samples (not shown)
gave values in the $30-34$~GHz/tesla range. These differences are
not fully understood. However, a possible explanation is the rather
small $\omega\tau$ product ($\sim 2$), and instrumental factors
which give rise to distorted CR lineshapes for some samples. We note
that the $\nu /B_{res}$ maximum is determined by the lowest-field
resonances. Therefore, although we correct distorted lineshapes (an
advantage to measuring phase shift in addition to
absorption\cite{MolaRSI}), a small systematic error may easily
account for a $10-15\%$ shift in A$_\circ$. Still, this is not
enough to account for the differences between this and earlier
studies. In fact, samples with the smallest A$_\circ$ values give
appreciably broader resonances (increase in $\tau$ of $30\%$)
relative to the samples A and~B used in this study, suggesting that
samples used in earlier studies may not have attained the fully
ordered metallic state. This raises the intriguing possibility that
perhaps A$_\circ$ and, therefore $\upsilon_F$, depends on cooling
rate, i.e. the closer proximity to the insulating state results in a
reduction in $\upsilon_F$, or larger effective mass.

Finally, we turn to the value of $\upsilon_F$. Since the zeros in
$\nu /B_{res}$ occur when the field is along $c^\prime$, the
appropriate value for $R_\parallel$ is 13.2~\AA.\cite{Ishiguro} This
gives rise to a Fermi velocity of $0.76(3)\times 10^5$~m/s. This is
within a factor of two of the value estimated from specific heat
measurements ($1.4\times 10^5$~m/s),\cite{Garoche82} and in close
agreement with the value estimated from recent transport
measurements ($3-6\times 10^4$~m/s).\cite{Kobayashi02}




In conclusion, we have observed periodic-orbit resonances for the
first time in a purely Q1D organic conductor. The behavior is
dominated by the $c^\prime$-axis dispersion. This technique provides
one of the most direct methods for determining key band structure
parameters which may be important for understanding the
unconventional superconductivity in the (TMTSF)$_2$X family.


We are grateful to A. E. Kovalev and J.S. Brooks for their
contributions to these studies. This work was supported by the
National Science Foundation (DMR0196461 and DMR0239481). S. H.
acknowledges the Research Corporation for financial support.


\clearpage

\noindent{{\bf Figure captions}}

\bigskip

\noindent{FIG. 1: Schematic of a Q1D FS ($\parallel b^\prime
c^*$-plane) with arbitrary warping. The angles and vectors are
described in the main text immediately after Eq.~1. Application of
a magnetic field causes electrons to traverse the FS in a plane
perpendicular to {\em B}. This reciprocal-space motion across the
FS corrugations results in a periodic modulation of the real space
velocity transverse to {\em a}, as represented by the wavy arrow.}

\bigskip

\noindent{FIG. 2: Angle dependent microwave absorption for sample~A
(2nd cooling) for rotations in the  $b^\prime c^*$-plane. Data were
obtained between $\theta_{b^\prime c^*}=-62.7^\circ$ and $7.6^\circ$
at $8.8^\circ$ intervals; the frequency was 61.8~GHz and the
temperature 2.5~K. The peaks in absorption are due to POR, and the
kink at high fields is due to the FISDW transition. Both features
are angle dependent.}

\bigskip

\noindent{FIG. 3: Angle dependence of the quantity $\nu/B_{res}$ for
(a) $b^\prime c^*$-plane rotations (Eq.~1a), and (b) rotations away
from the $b^\prime c^*$-plane towards {\em a} (Eq.~1b); the inserts
depict the experimental geometries. Data were obtained for two
samples (A and B); A1 and A2 denote successive cool downs for
sample~A. The solid curves are fits to Eq.~1, and the directions of
high symmetry are indicated in the figure. The experiments were all
performed at 2.5~K, and at frequencies ranging from 45 to 69~GHz
(not labeled). The quantity $\nu/B_{res}$ is frequency independent
to within the scatter of the data. In Fig.~1(a), the angle
dependence of $\nu/B_{\rm FISDW}$ is also shown (filled squares),
where $\nu=61.8$~GHz; these data have been fit to an inverse cosine
function (dashed curve), enabling an acurate determination of the
$b^\prime$ and $c^*$ directions.}

\clearpage

\begin{figure}

\includegraphics[width=0.45\textwidth]{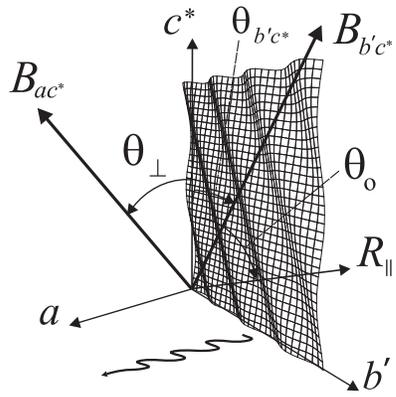}
\caption{\label{Fig1} S. Takahashi et al.}
\end{figure}
\clearpage

\begin{figure}
\includegraphics[width=0.45\textwidth]{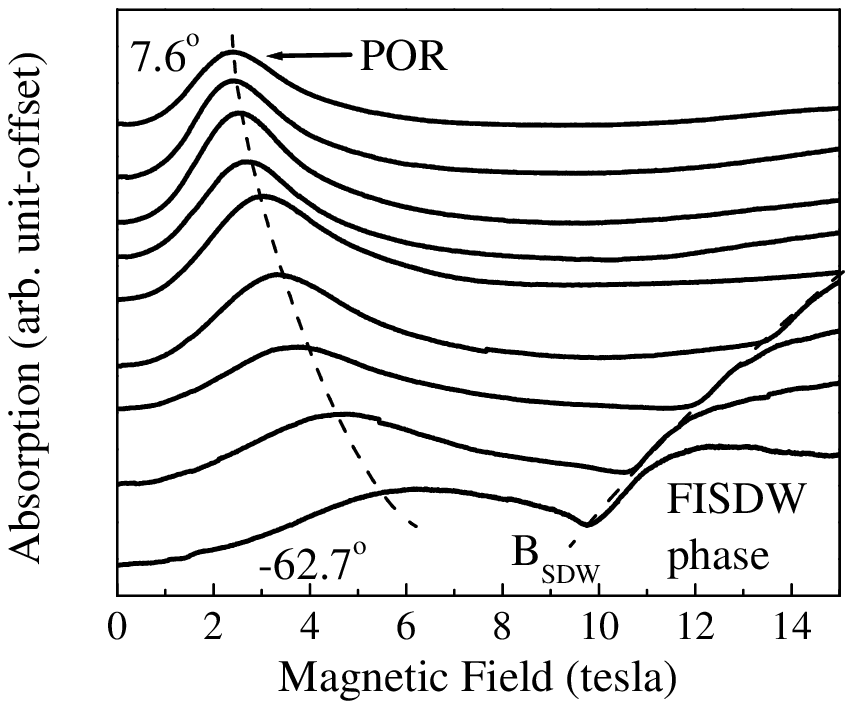}
\caption{\label{Fig2} S. Takahashi et al.}
\end{figure}

\begin{figure}
\includegraphics[width=0.45\textwidth]{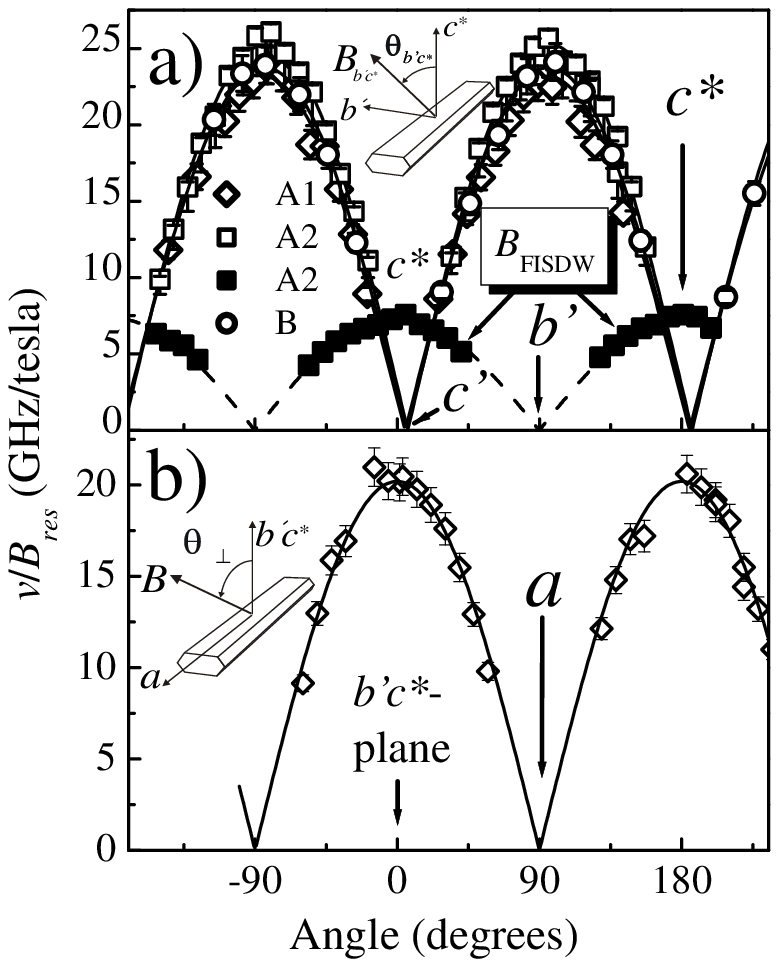}
\caption{\label{Fig3} S. Takahashi et al.}
\end{figure}



\end{document}